\documentclass[aps,prl,superscriptaddress,reprint]{revtex4-1}
\usepackage{amsmath}
\usepackage{amssymb}
\usepackage{graphicx}
\usepackage[colorlinks,urlcolor=blue]{hyperref}
\usepackage{url}
\usepackage{color,xcolor}
\usepackage{ulem}
\usepackage{float}
\usepackage{epstopdf}
\usepackage[none]{hyphenat}
\begin{document}

Notice: This manuscript has been authored by UT-Battelle, LLC, under contract DE-AC05-00OR22725 with the US Department of Energy (DOE). The US government retains and the publisher, by accepting the article for publication, acknowledges that the US government retains a nonexclusive, paid-up, irrevocable, worldwide license to publish or reproduce the published form of this manuscript, or allow others to do so, for US government purposes. DOE will provide public access to these results of federally sponsored research in accordance with the DOE Public Access Plan (https://www.energy.gov/doe-public-access-plan).
\title{General trends of electronic structures, superconducting pairing, and magnetic correlations in the Ruddlesden-Popper nickelate $m$-layered superconductors La$_{m+1}$Ni$_{m}$O$_{3m+1}$}
\author{Yang Zhang}
\author{Ling-Fang Lin}
\email{lflin@utk.edu}
\affiliation{Department of Physics and Astronomy, University of Tennessee, Knoxville, Tennessee 37996, USA}
\author{Adriana Moreo}
\affiliation{Department of Physics and Astronomy, University of Tennessee, Knoxville, Tennessee 37996, USA}
\affiliation{Materials Science and Technology Division, Oak Ridge National Laboratory, Oak Ridge, Tennessee 37831, USA}
\author{Satoshi Okamoto}
\affiliation{Materials Science and Technology Division, Oak Ridge National Laboratory, Oak Ridge, Tennessee 37831, USA}
\author{Thomas A. Maier}
\email{maierta@ornl.gov}
\affiliation{Computational Sciences and Engineering Division, Oak Ridge National Laboratory, Oak Ridge, Tennessee 37831, USA}
\author{Elbio Dagotto}
\email{edagotto@utk.edu}
\affiliation{Department of Physics and Astronomy, University of Tennessee, Knoxville, Tennessee 37996, USA}
\affiliation{Materials Science and Technology Division, Oak Ridge National Laboratory, Oak Ridge, Tennessee 37831, USA}
\date{\today}

\begin{abstract}
We report a comprehensive theoretical analysis of the Ruddlesden-Popper layered nickelates La$_{m+1}$Ni$_m$O$_{3m+1}$ ($m = 1$ to 6) under pressure.
These materials have recently received significant attention due to the discovery of superconductivity in some nickelates under pressure.
Our results suggest that, while these Ruddlesden-Popper layered nickelates display many similarities, they also show noticeable differences.
One of the common features of La$_{m+1}$Ni$_m$O$_{3m+1}$ is that the electronic states near the Fermi level are mainly contributed by Ni $3d$ orbitals,
slightly hybridized with O $2p$ orbitals.
The Ni $d_{3z^2-r^2}$ orbitals display bonding-antibonding, or bonding-antibonding-nonbonding, characteristic splittings, depending on the even or odd number of stacking layers $m$.
In addition, the ratio of the in-plane interorbital hopping between $d_{3z^2-r^2}$ and $d_{x^2-y^2}$ orbitals and in-plane intraorbital hopping between
$d_{x^2-y^2}$ orbitals was found to be large in La$_{m+1}$Ni$_m$O$_{3m+1}$ ($m = 1$ to 6), and this ratio increases from $m = 1$ to $m = 6$,
suggesting that the in-plane hybridization will increase as the layer number $m$ increases.
In contrast to the dominant $s^\pm$--wave state driven by spin fluctuations in the bilayer La$_3$Ni$_2$O$_7$ and trilayer La$_4$Ni$_3$O$_{10}$,
two nearly degenerate $d_{x^2-y^2}$-wave and $s^\pm$-wave leading states were obtained in the four-layer stacking La$_5$Ni$_4$O$_{13}$ and
five-layer stacking La$_6$Ni$_5$O$_{16}$. The leading $s^\pm$-wave state was recovered in the six-layer material La$_7$Ni$_6$O$_{19}$.
In general, at the level of the random phase approximation treatment, the superconducting transition temperature $T_c$ decreases in stoichiometric bulk systems
from the bilayer La$_3$Ni$_2$O$_7$ to the six-layer La$_7$Ni$_6$O$_{19}$, despite the $m$ dependent dominant pairing.
Both in-plane and out-of-plane magnetic correlations are found to be quite complex.
Within the in-plane direction, we obtained the peak of the magnetic susceptibility at ${\bf q} = (0.6 \pi, 0.6 \pi)$ for La$_5$Ni$_4$O$_{13}$ ($m = 4$)
and La$_7$Ni$_6$O$_{19}$ ($m=6$), and at ${\bf q} = (0.7 \pi, 0.7 \pi)$ for La$_6$Ni$_5$O$_{16}$ ($m = 5$).
Along the out-of-plane direction, four layers are coupled as  $\downarrow$\,--\,$\uparrow$\,--\,$\uparrow$\,--\,$\downarrow$ in La$_5$Ni$_4$O$_{13}$,
five layers are coupled as $\uparrow$\,--\,$\uparrow$\,--\,$\downarrow$\,--\,$\uparrow$\,--\,$\uparrow$ in La$_6$Ni$_5$O$_{16}$,
and six layers are coupled as $\uparrow$\,--\,$\downarrow$\,--\,$\downarrow$\,--\,$\uparrow$\,--\,$\uparrow$\,--\,$\downarrow$ in La$_7$Ni$_6$O$_{19}$.
\end{abstract}

\maketitle
\section{I. Introduction}
The discovery of superconductivity in infinite-layer Sr-doped NdNiO$_2$ ($d^9$ configuration) films with a $T_c$ of $\sim 15$ K~\cite{Li:Nature}
started the nickel era of unconventional high-temperature superconductivity~\cite{Nomura:Prb,Botana:prx,Li:prl20,Nomura:rpp,Zhang:prb20,Zeng:nc,Yang:nl,Gu:innovation},
which are reminiscent of the successful stories of the cuprates~\cite{Bednorz:Cu,Dagotto:rmp94} and iron-based superconductors~\cite{Kamihara:jacs,Dagotto:Rmp}.
Following the study of infinite-layer nickelates,  other NiO$_2$ layered materials were found to superconduct,
such as quintuple square-planar layered Nd$_6$Ni$_5$O$_{12}$ ($d^{\rm 8.8}$ configuration) with $T_c \sim 13$~K \cite{Pan:nm}.
Because their electron configuration is close to $d^9$, isoelectronic with Cu$^{2+}$,
these nickelates are expected to have a $d_{x^2-y^2}$-wave superconducting instability~\cite{Wu:prb,Sakakibara:prl} as in the cuprates~\cite{Dagotto:rmp94,Chaloupka2008}.
However, many theoretical and experimental efforts revealed fundamental differences between
square-planar layered nickelates and cuprates~\cite{Zhang:prb20,Jiang:prl,Werner:prb,Gu:prb,Karp:prx,Fowlie:np,Rossi:np}.
Very recently, pressure studies found superconductivity in bilayer La$_3$Ni$_2$O$_7$~\cite{Sun:arxiv} and trilayer La$_4$Ni$_3$O$_{10}$~\cite{Zhu:arxiv11,Li:cpl} nickelates with corner-shared NiO$_6$ octahedra layers.
This rapidly developing new avenue for the study of nickelate superconductors has already attracted much attention in Condensed Matter Physics and Material Sciences~\cite{LiuZhe:arxiv,Zhang:arxiv-exp,Hou:arxiv,Yang:arxiv09,Wang:arxiv9,Dong:arxiv12,Sakakibara:arxiv09,Zhang:prb24,Zhang:2323,Zhang:arxiv25,Zhang:Z2-prl,Xie:SB,Chen:arxiv2024,Dan:arxiv2024,Takegami:prb,Xu:prb25,Chen:prl24,Wang:nature}.

\begin{figure*}
\centering
\includegraphics[width=0.92\textwidth]{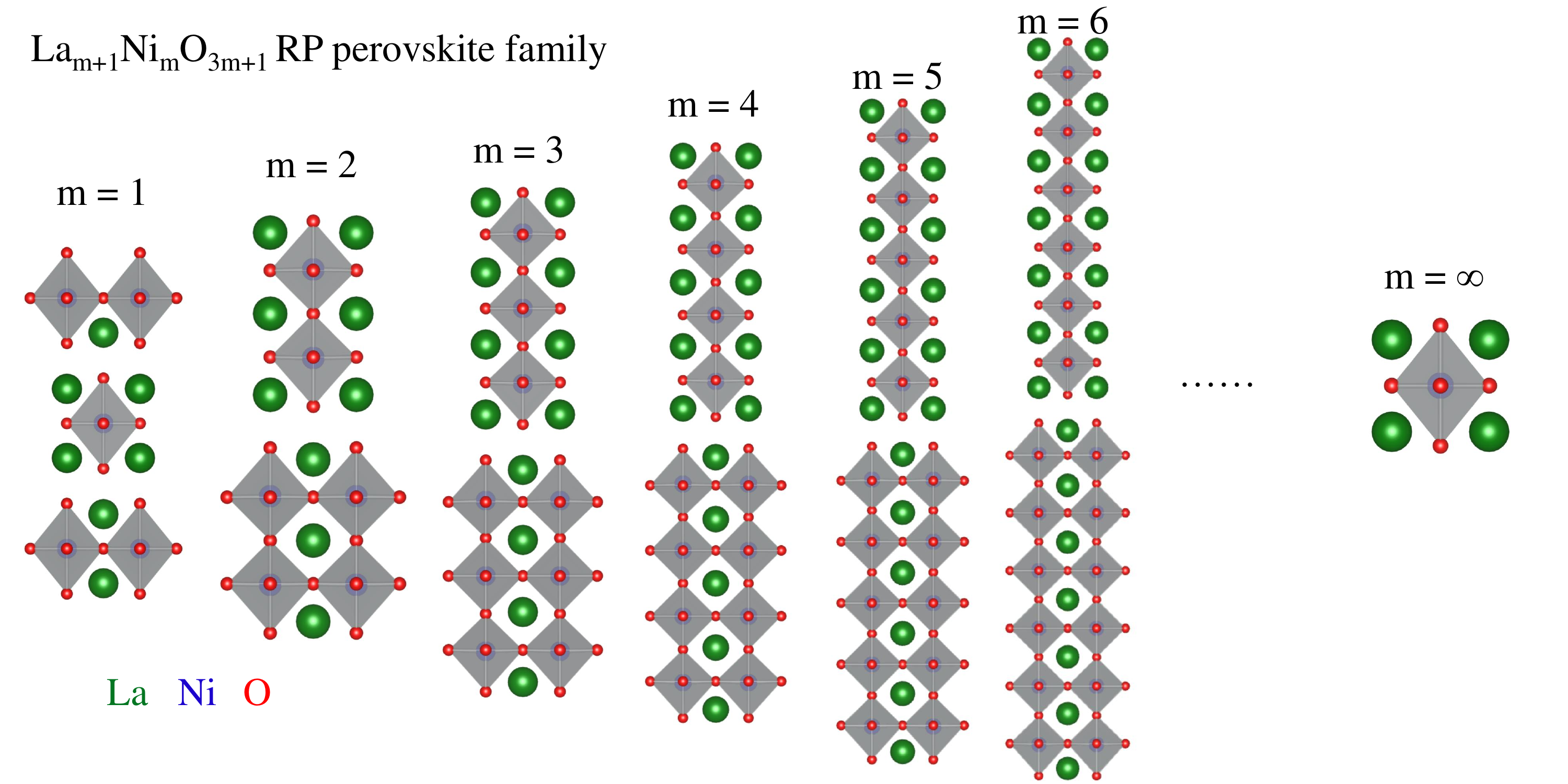}
\caption{ Schematic crystal structures of the RP perovskite family La$_{m+1}$Ni$_m$O$_{3m+1}$ (green = La; blue = Ni; red = O). The visualization code VESTA is used~\cite{Momma:vesta}.}
\label{crystal}
\end{figure*}

La$_3$Ni$_2$O$_7$ and La$_4$Ni$_3$O$_{10}$ belong to the Ruddlesden-Popper (RP) perovskite family La$_{ m+1}$Ni$_m$O$_{3m+1}$ ($m = 1$ to $\infty$) involving bilayer and trilayer blocks stacking structures, respectively, as displayed in Fig.~\ref{crystal}.
In contrast to the square-planar layered nickelates La$_{m+1}$Ni$_m$O$_{2m+2}$ ($m = 1$ to $\infty$)~\cite{Lacorre:jssc,LaBollita:prb21,Jung:prb22}, there are  apical oxygens connecting the Ni layers in the RP nickelate family La$_{m+1}$Ni$_m$O$_{3m+1}$ ($m = 1$ to $\infty$). At ambient pressure, bilayer La$_3$Ni$_2$O$_7$ has an Amam (No. 63) structure~\cite{Ling:jssc}, and thus, the NiO$_6$ octahedra are distorted. By applying hydrostatic pressure, the distortion of NiO$_6$ octahedra is suppressed around 14 GPa~\cite{Zhang:arxiv-exp}, accompanied by a first-order structural phase transition,
resulting in an Fmmm (No. 69) phase without a tilting distortion of the NiO$_6$ octahedra~\cite{Sun:arxiv}.
By further increasing pressure, a superconducting state appears in a very broad pressure regime~\cite{Sun:arxiv,Li:arxiv24}.  The initial claim of superconductivity in the
La$_3$Ni$_2$O$_7$ system was based on measurements of the resistance by a four-terminal device, where the presence of zero resistance and the Meissner effect~\cite{Sun:arxiv} was deduced by observing a sharp transition and a flat stage in resistance when using potassium bromide as the pressure-transmitting medium and a diamagnetic response in the susceptibility.
Later, actual zero resistance~\cite{Hou:arxiv,Zhang:arxiv-exp,Zhang:jmst} and Meissner effect~\cite{Li:arxiv24} were observed by several groups, establishing the superconducting state in bilayer La$_3$Ni$_2$O$_7$.
The superconducting volume fraction was reported to be around $48 \%$ in their high-quality superconducting samples by using the $ac$ magnetic susceptibility~\cite{Li:arxiv24}. Very recently, a tetragonal I4/mmm (No. 139) phase was proposed to be the correct high-pressure structure of La$_3$Ni$_2$O$_7$ from both theory~\cite{Geisler:qm} and experiment~\cite{Wang:jacs}, instead of the Fmmm symmetry. The tilting distortion of the NiO$_6$ octahedra is suppressed in both phases. Due to the very small distortion from I4/mmm (No. 139) to Fmmm (No. 69) in La$_3$Ni$_2$O$_7$~\cite{Zhang:1313}, those two phases are expected to display no fundamental differences, providing the same physics~\cite{Sakakibara:prl24}.

As in the bilayer La$_3$Ni$_2$O$_7$, the NiO$_6$ octahedra are also distorted in the trilayer nickelate La$_4$Ni$_3$O$_{10}$ with a $P21/c$ (No. 14) phase~\cite{Zhang:nc20,Li:nc,Puggioni:prb,Zhang:prm,Rout:prb} at ambient pressure. Under hydrostatic pressure, the distortion of the NiO$_6$ octahedra is suppressed, and a high-symmetry I4/mmm phase appears at around 15 GPa~\cite{Zhu:arxiv11,Li:cpl,Leonov:arxiv12}, once again without the tilting of oxygen octahedra. Following this structural transition, the superconducting state was observed in a broad pressure range with a $T_c$ of about 20--30~K~\cite{Sakakibara:arxiv09,Li:cpl,Zhu:arxiv11,Zhang:arxiv11}, which is lower than in the bilayer La$_3$Ni$_2$O$_7$~\cite{Sun:arxiv}. The Meissner effect was also observed in {\it ac} magnetic susceptibility experiments with superconducting volume fraction around $\sim 80 \%$ ~\cite{Li:cpl}.
However, the single-layer RP nickelate La$_2$NiO$_4$ (see Fig.~\ref{crystal}) has never been reported to superconduct at both ambient conditions and high pressure~\cite{Zhang:jmst}.

Theoretical studies have played a crucial role in providing insights into the nature of superconducting states in nickelates~\cite{Zhang:nc24,Christiansson:arxiv,Yang:arxiv,Shen:arxiv,Liu:arxiv,Yang:arxiv1,Oh:arxiv,Liao:arxiv,Cao:arxiv,Lechermann:arxiv,Shilenko:arxiv,Jiang:arxiv,Huang:arxiv,Zhang:prb23-2,Qin:arxiv,Wang:arxiv12,Li:scpma,Wang:PRB24,
Zhang:PRB24,Huang:PRB24,Lin:prb23,Oh:prb24,Zhang:prb25,Chen:prb24,Ryee:prl,Heier:prb,Zhan:prl,Geisler:qm24,LaBollita:prm,Lu:prb25,Duan:arxiv25,Braz:arxiv25,Yang:prb25}. Density functional theory (DFT) calculations suggest that both La$_3$Ni$_2$O$_7$ and La$_4$Ni$_3$O$_{10}$ nickelates systems
can be described using a two-$e_g$-orbital model in a bilayer or in a trilayer Ni lattice~\cite{Sakakibara:arxiv09,Luo:prl23,Zhang:prb23,Zhang:arxiv24}. It is also indicated that the Fermi surfaces are contributed by Ni $d_{x^2-y^2}$ and $d_{3z^2-r^2}$ orbitals~\cite{Luo:prl23,Zhang:prb23,Zhang:arxiv24,Li:nc,LaBollita:prb24}, with two-electron sheets with mixed $e_g$ orbitals and a small hole pocket from the Ni $d_{3z^2-r^2}$ orbital at high pressure which is absent at ambient pressure. Angle-resolved photoemission spectroscopy experiments confirmed the Fermi surface obtained by DFT calculations for both La$_3$Ni$_2$O$_7$ and La$_4$Ni$_3$O$_{10}$ at ambient pressure~\cite{Yang:arxiv09,Li:nc}.

Based on the two-$e_g$-orbitals model, theoretical studies suggested that $s_{\pm}$-wave pairing is dominant in bilayer La$_3$Ni$_2$O$_7$, induced by spin fluctuations due to the partial nesting of the Fermi surfaces with wave vectors approximately ($\pi$, 0) or (0, $ \pi$)~\cite{Yang:arxiv,Zhang:nc24,Liu:arxiv,Liao:arxiv,Qu:prl,Lu:prl,Tian:prb24}. The $s_{\pm}$-wave pairing channel is believed to be driven mainly by strong interlayer coupling via the $d_{3z^2-r^2}$ orbital, but the $d_{x^2-y^2}$ orbital also contributes robustly to the superconducting gap functions, comparable in some cases to the contributions of the $d_{3z^2-r^2}$ orbital~\cite{Tian:prb24,Zhang:prb23-2}. Alternatively, there also appeared other theoretical studies, suggesting a dominant $d_{x^2-y^2}$-wave or  $d_{xy}$-wave superconducting pairing channel driven by the intralayer coupling~\cite{Lechermann:arxiv,Jiang:arxiv,Liu:arxiv2023,Fan:arxiv23}. In fact, the interplay between the intralayer and the interlayer pairing tendency was discussed many years ago by using $t-J$ models, where a prevailing interlayer coupling leads to $s$-wave pairing while a prevailing stronger intralayer coupling results in $d$-wave pairing~\cite{Dagotto:prb92}.

Considering these established results for bilayer La$_3$Ni$_2$O$_7$ and trilayer La$_4$Ni$_3$O$_{10}$, it is natural to wonder about the properties of other members of the RP nickelates family La$_{m+1}$Ni$_m$O$_{3m+1}$:
Could the RP higher-order $m = 4-6$ stacking nickelate be superconducting under pressure? In this case, what trends of superconductivity dominate in La$_{m+1}$Ni$_m$O$_{3m+1}$? What is the dependence of the pairing channel under pressure in La$_{m+1}$Ni$_m$O$_{3m+1}$? What are the magnetic correlations in La$_{m+1}$Ni$_m$O$_{3m+1}$? Could we establish a general intuitive picture of the RP nickelates systems?

To address these questions, here we theoretically study the high-symmetry I4/mmm phase of RP nickelates La$_{m+1}$Ni$_{m}$O$_{3m+1}$ ($m = 1$ to 6) without the tilting of the NiO$_6$ octahedra under pressure by using first-principles DFT and random phase approximation (RPA) calculations.
Our studies find many similarities but also some differences in the RP nickelates with different $m$-layer blocks. The $d_{3z^2-r^2}$ orbital displays the bonding-antibonding, or the bonding-antibonding-nonbonding splitting character,
depending on the even or odd number of stacking layers $m$. In addition, the in-plane ratio of the interorbital hopping between the $e_g$ orbitals and intraorbital hopping between $d_{x^2-y^2}$ orbitals was found to be large in La$_{m+1}$Ni$_{m}$O$_{3m+1}$,
and this ratio increases from $m = 1$ to $m = 6$.  In contrast to the $s^\pm$-wave pairing driven by spin fluctuations in the bilayer La$_3$Ni$_2$O$_7$ and trilayer La$_4$Ni$_3$O$_{10}$, two nearly degenerate $d_{x^2-y^2}$-wave and $s^\pm$-wave instabilities are obtained
in La$_5$Ni$_4$O$_{13}$ and La$_6$Ni$_5$O$_{16}$, while a leading $s^\pm$-wave state was found in La$_7$Ni$_6$O$_{19}$. In general, at the level of the RPA treatment, the superconducting transition temperature $T_c$ was expected to decrease in stoichiometric bulk systems from the bilayer La$_3$Ni$_2$O$_7$ to the six-layer La$_7$Ni$_6$O$_{19}$. Moreover, in the Ni planes, we obtained the peak of the magnetic susceptibility at ${\bf q}=(0.6\pi, 0.6\pi)$ for La$_5$Ni$_4$O$_{13}$ and La$_7$Ni$_6$O$_{19}$, and at ${\bf q}=(0.7\pi, 0.7\pi)$ for La$_6$Ni$_5$O$_{16}$.
Furthermore, the out-of-plane magnetic correlations are complex in La$_{m+1}$Ni$_{m}$O$_{3m+1}$, with details depending on the stacking layer number $m$.

The scope of this work is to explore and understand the trends in the RP nickelates family La$_{m+1}$Ni$_{m}$O$_{3m+1}$ with different $m$-layer stacking. Considering previous studies of superconductivity in the bilayer La$_3$Ni$_2$O$_7$ and trilayer La$_4$Ni$_3$O$_{10}$, here
we use the high-symmetry I4/mmm structure (see Fig.~\ref{crystal}) without the tilting of NiO$_6$ octahedra to perform our study. The paper is organized as follows: in Section II the methods used are described, the results are presented in Section III, and Section IV is devoted to the conclusions and discussions.

\section{II. Method}
In this work, first-principles DFT calculations were performed employing the Vienna {\it ab initio} simulation package (VASP) code by using the projector augmented wave  method~\cite{Kresse:Prb,Kresse:Prb96,Blochl:Prb} with
the generalized gradient approximation and the Perdew-Burke-Ernzerhof (PBE) exchange potential~\cite{Perdew:Prl}. The plane-wave cutoff energy was set as $550$~eV. Both lattice constants and atomic positions
were fully relaxed until the Hellman-Feynman force on each atom was smaller than $0.01$ eV/{\AA}. In addition, the $\bf k$-point mesh used was $20\times20\times2$ for the I4/mmm phase of La$_{m+1}$Ni$_{m}$O$_{3m+1}$ at 30 GPa.

To carry out further theoretical analyses, we derive maximally localized Wannier functions (MLWFs) projected on Ni $3d$ states using the WANNIER90 package~\cite{Mostofi:cpc}
and construct low-energy $e_g$-orbital tight-binding models, consisting of orbital-dependent hopping matrices and crystal-field splitting.
The resulting tight-binding Hamiltonian is expressed as
\begin{eqnarray}
\label{eq:Hk}
H_k = \sum_{\substack{i \gamma\gamma' \\\vec{\alpha}\sigma}}t_{\gamma\gamma'}^{\vec{\alpha}}
\bigl(c^{\dagger}_{i \gamma\sigma}c^{\phantom\dagger}_{i+\vec{\alpha} \gamma'\sigma}+H.c.\bigr)+ \sum_{i \gamma\sigma} \Delta_{\gamma} n_{i\gamma\sigma}\,.
\end{eqnarray}
The first term represents the hopping of an electron from orbital $\gamma'$ at site $i+\vec{\alpha}$ to orbital $\gamma$ at site $i$.
$c^{\dagger}_{i\gamma\sigma}$ ($c^{\phantom\dagger}_{i\gamma\sigma}$) is the creation (annihilation) operator of an electron at site $i$, orbital $\gamma$ with spin $\sigma$.
$\Delta_{\gamma}$ represents the crystal field of orbital $\gamma$ with $n_{i \gamma \sigma }=c_{i \gamma \sigma}^\dag c_{i \gamma \sigma}$. The vectors $\vec{\alpha}$ are along the three directions, defining different hopping neighbors. In addition, the detailed input files of hopping matrices and crystal-field splittings can be found in a separate attachment in the Supplemental Materials~\cite{Supplemental}. Based on the obtained Fermi energy for the stoichiometric ratio filling of La$_{m+1}$Ni$_{m}$O$_{3m+1}$ ($m = 1$ to 6) in our tight-binding model calculations, a $4001\times4001$ $\bf k$-mesh was used to calculate the Fermi surface.

To discuss the superconducting pairing and magnetic correlations in different $m$-layer stacking nickelates under pressure, here we used the many-body RPA method based on a perturbative weak-coupling expansion in the Hubbard interaction, similar to our previous analysis~\cite{Zhang:nc24,Zhang:arxiv24}.
We considered a multiorbital Hubbard model, including the kinetic energy $H_{\rm k}$ and local interaction $H_{\rm int}$ terms.
The model is written as $H = H_{\rm k} + H_{\rm int}$,
where $H_{\rm int}$ includes the intraorbital Hubbard repulsion $U$, the interorbital Hubbard repulsion $U'$, the Hund's coupling $J$, and the interorbital electron-pair hopping $J'$, given by
\begin{eqnarray}
\label{eq:Hint}
H_{\rm int}= U\sum_{i\gamma}n_{i \gamma\uparrow} n_{i \gamma\downarrow} +
\biggl(U'-\frac{J}{2}\biggr)\sum_{\substack{i\\\gamma < \gamma'}} n_{i \gamma} n_{i\gamma'} \nonumber \\
-2J \sum_{\substack{i\\\gamma < \gamma'}} {{\bf S}_{i \gamma}}\cdot{{\bf S}_{i \gamma'}}+J' \sum_{\substack{i\\\gamma < \gamma'}} \bigl(P^{\dagger}_{i\gamma} P^{\phantom{\dagger}}_{i\gamma'}+H.c.\bigr).
\end{eqnarray}
Here, the standard relations $U'=U-2J$ and $J' = J$ are assumed, with $n_{i \gamma} = n_{i \gamma \uparrow}+ n_{i \gamma \downarrow}$ and $P_{i\gamma}$=$c_{i\gamma\downarrow} c_{i\gamma \uparrow}$.

In the multi-orbital RPA approach \cite{Kubo2007,Graser2009,Altmeyer2016,Romer2020,Mishra:sr,Maier:prb22}, the spin susceptibility is obtained from the bare susceptibility (Lindhart function) $\chi_0({\bf q})$ as $\chi({\bf q}) = \chi_0({\bf q})[1-{\cal U}\chi_0({\bf q})]^{-1}$. Here, $\chi_0({\bf q})$ is an orbital-dependent susceptibility tensor and ${\cal U}$ is a tensor that contains the intra-orbital $U$ and inter-orbital $U'$ density-density interactions, the Hund's rule coupling $J$, and the pair-hopping $J'$ term. The pairing strength $\lambda_\alpha$ for channel $\alpha$ and the corresponding gap structure $g_\alpha({\bf k})$ are obtained from solving an eigenvalue problem of the form
\begin{eqnarray}\label{eq:pp}
	\int_{FS} d{\bf k'} \, \Gamma({\bf k -  k'}) g_\alpha({\bf k'}) = \lambda_\alpha g_\alpha({\bf k})\,,
\end{eqnarray}
where the momenta ${\bf k}$ and ${\bf k'}$ are on the Fermi Surface FS and $\Gamma({\bf k - k'})$ contains the irreducible particle-particle vertex. While both spin and charge susceptibilities contribute to the pairing interaction, the dominant contribution comes from the RPA spin susceptibility tensor $\chi^s_{\ell_1\ell_2\ell_3\ell_4}({\bf k-k'})$, where $\{\ell_i\}$ denote the different $e_g$ orbitals of the RP nickelates, as shown in Fig.~\ref{crystal}. For the electronic densities, we use the stoichiometric cases (i.e. 5 electrons for $m = 4$, corresponding to 1.25 electrons per site for the two $e_g$ orbitals).

\section{III. Results}

\subsection{A. Model systems}
\begin{figure*}
\centering
\includegraphics[width=0.92\textwidth]{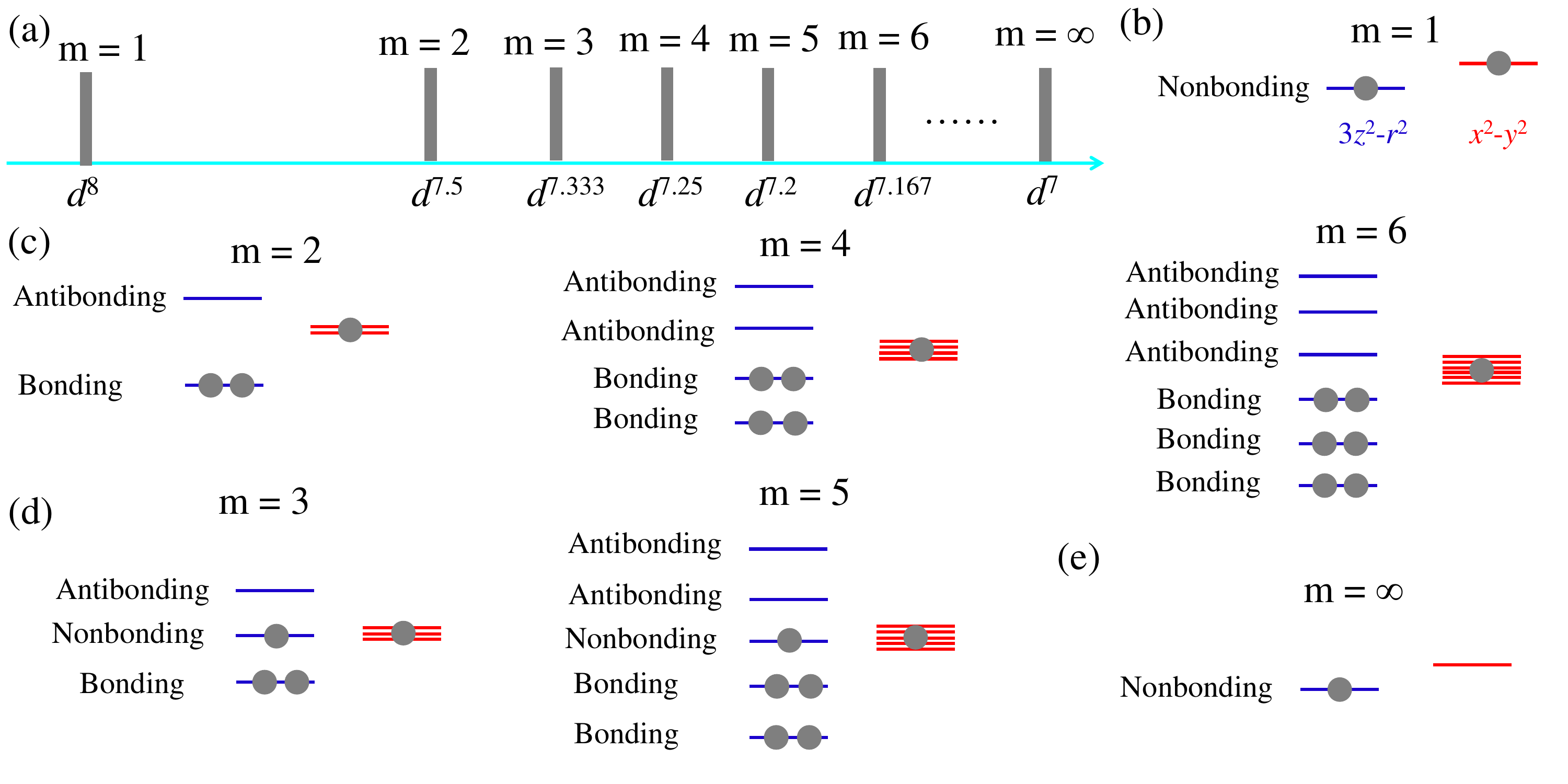}
\caption{(a) Schematic electronic densities of $3d$-electrons per Ni for different $m$-layer stacking RP nickelates La$_{m+1}$Ni$_m$O$_{3m+1}$. (b-e) Sketches of electronic states for two ``active'' $e_g$ orbitals in La$_{m+1}$Ni$_m$O$_{3m+1}$ with layers (b) $m = 1$, (c) $m = 2, 4, 6$, and (d) $m = 3, 5$ and (e) $\infty$, respectively. The light blue and pink horizontal lines represent $d_{3z^2-r^2}$ and $d_{x^2-y^2}$ states, respectively. The solid circles indicate the occupied electrons of $e_g$ orbitals in the stoichiometric ratio La$_{m+1}$Ni$_{m}$O$_{3m+1}$.}
\label{Electronic-density}
\end{figure*}

RP nickelates La$_{m+1}$Ni$_{m}$O$_{3m+1}$ ($m$ from 1 to $\infty$) have $m$-layer blocks of corner-shared NiO$_6$ octahedra.
Considering the formal valence of La$^{3+}$ and O$^{2-}$, the average valence of Ni in RP nickelates is given by $3-1/m$ as a function of $m$,
leading to the $3d$ electron density $n$ varying from 8 at $m = 1$ to 7 at $m = \infty$ as shown in Fig.~\ref{Electronic-density}(a).
Since Ni $t_{2g}$ orbitals are fully occupied, these nickelates could be regarded as ``effective'' two-$e_g$-orbital systems.

An important consequence of $m$-layer stacking in RP nickelates is the bonding-antibonding splitting of Ni $d_{3z^2-r^2}$ orbitals.
Due to the single-layer blocks stacking, $d_{3z^2-r^2}$ orbitals do not have bonding-antibonding splitting in La$_2$NiO$_4$ ($m = 1$ case), as shown in Fig.~\ref{Electronic-density}(b).
For the multiple layer stacking geometries in La$_{m+1}$Ni$_{m}$O$_{3m+1}$, the $d_{3z^2-r^2}$ orbital shows the bonding-antibonding, or the bonding-antibonding-nonbonding splitting, depending on the even or odd number of stacking layers.
Similar to the ``dimer'' physics in the $m = 2$ RP nickelate, the $m = 4$ and $m = 6$ compounds show an orbital-selective behavior~\cite{Streltsovt:prb14,Zhang:ossp}, where the $d_{3z^2-r^2}$ orbital splits into bonding and antibonding states,
while the $d_{x^2-y^2}$ orbital remains decoupled among planes, as shown in Fig.~\ref{Electronic-density} (c). Similar to the ``trimer'' physics in the $m = 3$ RP nickelate, the $d_{3z^2-r^2}$ orbitals have bonding-antibonding and nonbonding
splitting for La$_6$Ni$_5$O$_{16}$ ($m = 5$ case), as displayed in Fig.~\ref{Electronic-density} (d). At the other limit ($m = \infty$), i.e. the cubic perovskite LaNiO$_3$, the $d_{3z^2-r^2}$ orbital does not have bonding-antibonding splitting
as well (see Fig.~\ref{Electronic-density}(e)). However, due to the strong in-plane hybridization between the $e_g$ orbitals, both $d_{3z^2-r^2}$ and $d_{x^2-y^2}$ orbitals are not integer-filled, leading to a ``self-doping'' effect
in those two orbitals. Thus, this effect is also expected in all other $m$-layer stackings La$_{m+1}$Ni$_{m}$O$_{3m+1}$.

Note that the scope of this work is to explore and understand the trends in the RP nickelates family La$_{m+1}$Ni$_{m}$O$_{3m+1}$ with different $m$-layer stacking. Considering previous studies on the bilayer and trilayer nickelate bulk superconductors where superconductivity was found in the high-symmetry phase under pressure where the tilting of NiO$_6$ octahedra is absent, here we use the high-symmetry I4/mmm structure~\cite{Jung:prb22} (see Fig.~\ref{crystal}) to mimic the uniform lattice structure expected for RP nickelates at 30 GPa. Since the electronic states near the Fermi level of RP nickelates La$_{m+1}$Ni$_m$O$_{3m+1}$ ($m = 1$ to 6) are mainly contributed by Ni $3d$ orbitals while the O $2p$ orbitals are located deeper in energy, a large charge-transfer gap between Ni $3d$ and O $2p$ states is expected as a common feature in these systems.

\subsection{B. Tight-binding band structures and Fermi surfaces}

To explore and understand the trends in the RP nickelates family with different $m$-layer stacking, we considered two $e_g$-orbitals for each $m$-layer system, i.e., the model has $2m$ bands. Based on the optimized crystal structures at 30 GPa, we obtained hoppings and crystal-field splittings by using the MLWFs method in the WANNIER90 packages~\cite{Mostofi:cpc}, and constructed an ${e_g}$-orbital tight-binding model for La$_{m+1}$Ni$_m$O$_{3m+1}$ ($m$ = 1 to 6).

\begin{figure}
\centering
\includegraphics[width=0.48\textwidth]{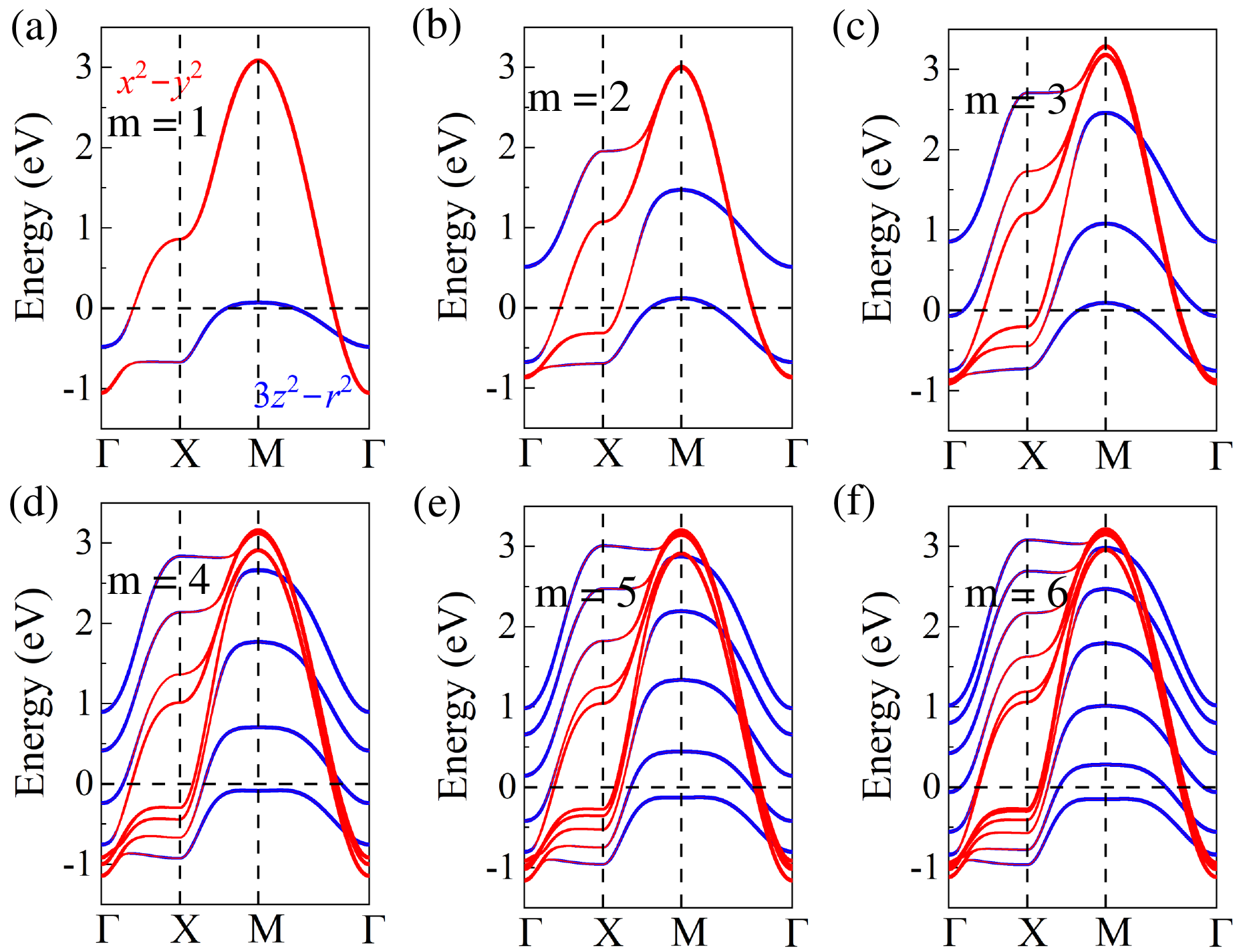}
\caption{ Band structures of the tight-binding models for La$_{m+1}$Ni$_m$O$_{3m+1}$ ($m$ = 1 to 6) at 30 GPa. The coordinates of the high symmetry points in the Brillouin zone are $\Gamma$ = (0, 0, 0), X = (0, 0.5, 0), and M = (0.5, 0.5, 0). Here, two $e_g$ orbitals were considered in the tight-binding models with an overall filling of $n = 2$ to 7 for $m = 1$ to 6 (e.g., 1.2 electrons per site for $m = 6$).}
\label{TB}
\end{figure}

In Fig.~\ref{TB}, we plot the tight-binding band structures of the I4/mmm phase for different $m$-layer stacking at 30 GPa. By increasing $m$ from 1 to 6, the bandwidth of $e_g$ orbitals is slightly increased by about $3 \%$.
In bilayer La$_3$Ni$_2$O$_7$ and trilayer La$_4$Ni$_3$O$_{10}$, a small hole pocket was found at the M point due to the bonding $d_{3z^2-r^2}$ orbitals. Such a small hole pocket is absent in La$_{m+1}$Ni$_m$O$_{3m+1}$ with $m$ = 4 to 6, because the lowest bonding $d_{3z^2-r^2}$ orbital does not touch the Fermi level.

\begin{figure}
\centering
\includegraphics[width=0.48\textwidth]{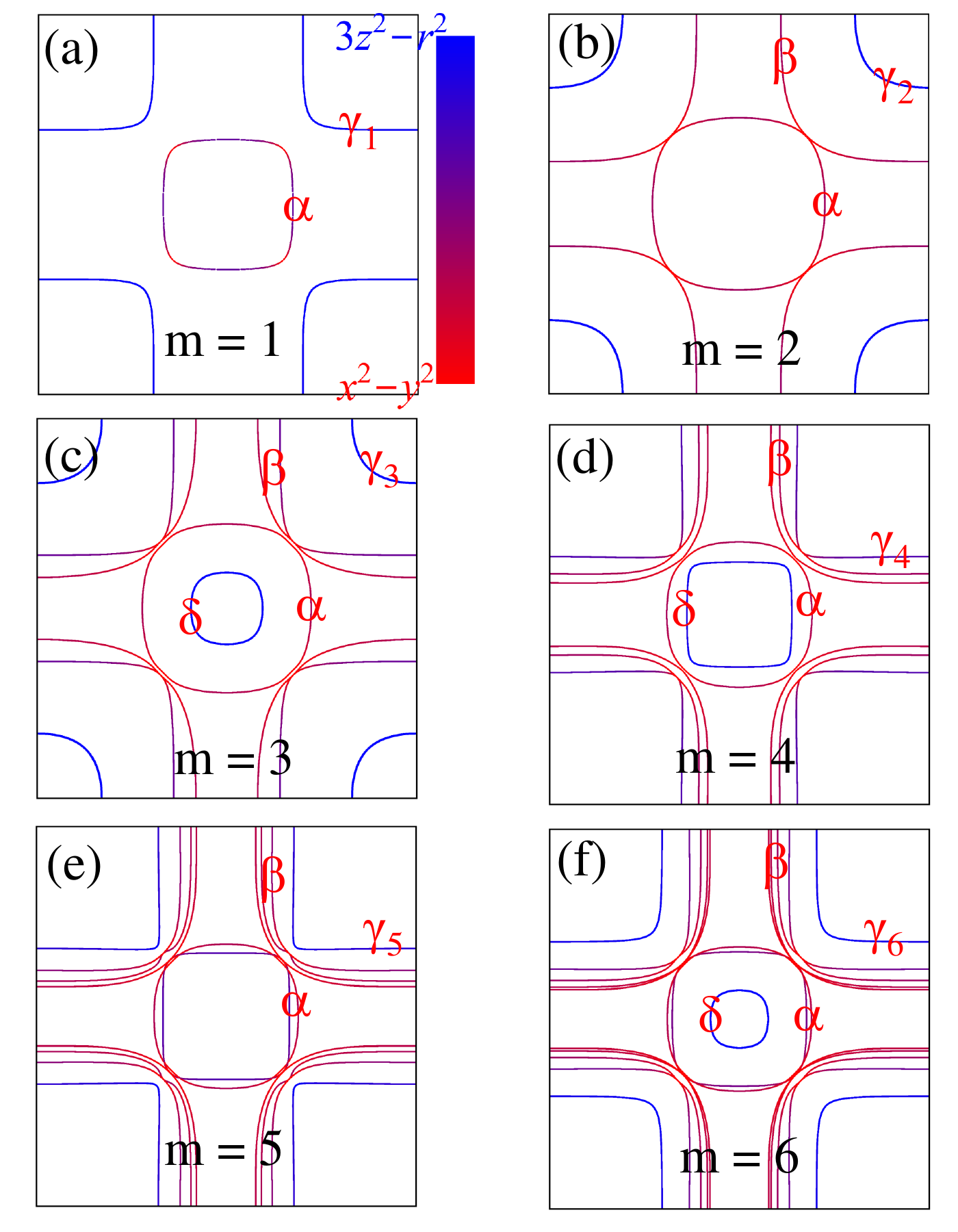}
\caption{ Fermi surfaces of the tight-binding models for La$_{m+1}$Ni$_m$O$_{3m+1}$ ($m$ = 1 to 6) at 30 GPa. Here, two $e_g$ orbitals were considered in the tight-binding models with an overall filling of $n = 2$ to 7 for $m = 1$ to 6 (e.g., 1.2 electrons per site for $m = 6$).}
\label{FS}
\end{figure}

Figure~\ref{FS} summarizes the Fermi surfaces of RP nickelates. As shown in Fig.~\ref{FS}(a), there are two bands crossing the Fermi level in the single-layer La$_2$NiO$_4$ at 30 GPa, leading to an electronic sheet made of the mixed $d_{3z^2-r^2}$ and $d_{x^2-y^2}$ orbitals and a hole sheet originating from the $d_{3z^2-r^2}$ orbital. As mentioned above, La$_3$Ni$_2$O$_7$ and La$_4$Ni$_3$O$_{10}$ have small hole pockets around the M point (see Figs.\ref{FS}(b) and (c)), and such a hole pocket is missing in La$_{m+1}$Ni$_m$O$_{3m+1}$ ($m$ = 4 to 6) at 30 GPa as displayed in Figs.~\ref{FS}~(d-f). In addition, a hole sheet from the $d_{3z^2-r^2}$ orbital was found near the M point in La$_6$Ni$_5$O$_{16}$ (see $\gamma_5$ in Fig.\ref{FS}(e)) and La$_7$Ni$_6$O$_{19}$ (see $\gamma_6$ in Fig.\ref{FS}(f)). Similar to the trilayer La$_4$Ni$_3$O$_{10}$ case~\cite{Zhang:arxiv24}, an electron $\sigma$ pocket made up of a $d_{3z^2-r^2}$ orbital was also also obtained in La$_7$Ni$_6$O$_{19}$, as in Fig.\ref{FS}(f). The tight-binding Fermi surfaces of the $m = 4$ to $m = 6$ cases discussed here are also in agreement with previous DFT calculations in the tetragonal I4/mmm phase of high-order RP nickelates~\cite{Jung:prb22}.

We also notice the characteristic trend of hopping intensities. In the single-layer La$_2$NiO$_4$ ($\sim 0.333$) at 30 GPa, the in-plane hybridization ($t_{x/y}^{12}$/$t_{x/y}^{22}$) between the $d_{3z^2-r^2}$ and $d_{x^2-y^2}$ orbitals is significantly small, compared to the bilayer and trilayer nickelates ($\sim 0.475$ for La$_3$Ni$_2$O$_7$ and $\sim 0.532/0.543$ for La$_4$Ni$_3$O$_{10}$)~\cite{Zhang:nc24,Zhang:arxiv24}. This value $t_{x/y}^{12}$/$t_{x/y}^{22}$ continues to  increase with the increase of high-order layer stacking at 30 GPa: $\sim 0.543/0.564$ for La$_5$Ni$_4$O$_{13}$, $\sim 0.545/0.573/0.580$ for La$_6$Ni$_5$O$_{16}$, and $\sim 0.544/0.568/0.584$ for La$_7$Ni$_6$O$_{19}$, suggesting that the in-plane hybridization between the $d_{3z^2-r^2}$ and $d_{x^2-y^2}$ orbitals and
the layer number $m$ has a positive correlation.

\subsection{C. Superconducting pairing tendency}
Next, we study the superconducting pairing tendencies in the RP nickelates. Here, we carry out the RPA technique to investigate the pairing instability of the multi-orbital Hubbard model, $H=H_k+H_{int}$, where $H_k$ is the single-particle part of the Hamiltonian with hopping matrix and crystal-field splittings from the I4/mmm phase of La$_{m+1}$Ni$_m$O$_{3m+1}$ at 30 GPa, given by Eq.~(\ref{eq:Hk}). The RPA calculations of the pairing vertex are based on a perturbative weak-coupling expansion with respect to the electron-electron interaction part of the Hamiltonian (see Method section), $H_{int}$, given by Eq.~(\ref{eq:Hint}), where the local Coulomb interaction matrix involves the intra-orbital ($U$), inter-orbital ($U'$), Hund's rule coupling ($J$), and pair-hopping ($J'$) terms~\cite{Kubo2007,Graser2009,Altmeyer2016,Romer2020}.

As discussed in previous work, the bilayer La$_3$Ni$_2$O$_7$~\cite{Zhang:nc24} and trilayer La$_4$Ni$_3$O$_{10}$
~\cite{Zhang:arxiv24} have leading $s^\pm$-wave pairing instabilities driven by spin-fluctuations   with different nesting vectors: ${\bf q} =  (\pi, 0)$ or $(0, \pi)$ for La$_3$Ni$_2$O$_7$ and ${\bf q} = (\pi, \pi)$ for La$_4$Ni$_3$O$_{10}$.
For the high-order nickelates ($m = 4-6$), starting from the high-symmetry I4/mmm phase obtained from the previous study~\cite{Jung:prb22}, we fully relaxed the atomic positions and lattice constants at 30 GPa.
In the following, we focus on La$_{m+1}$Ni$_m$O$_{3m+1}$ with $m = 4$ to 6 and explore their  superconducting pairing tendencies.

\begin{figure}
\centering
\includegraphics[width=0.48\textwidth]{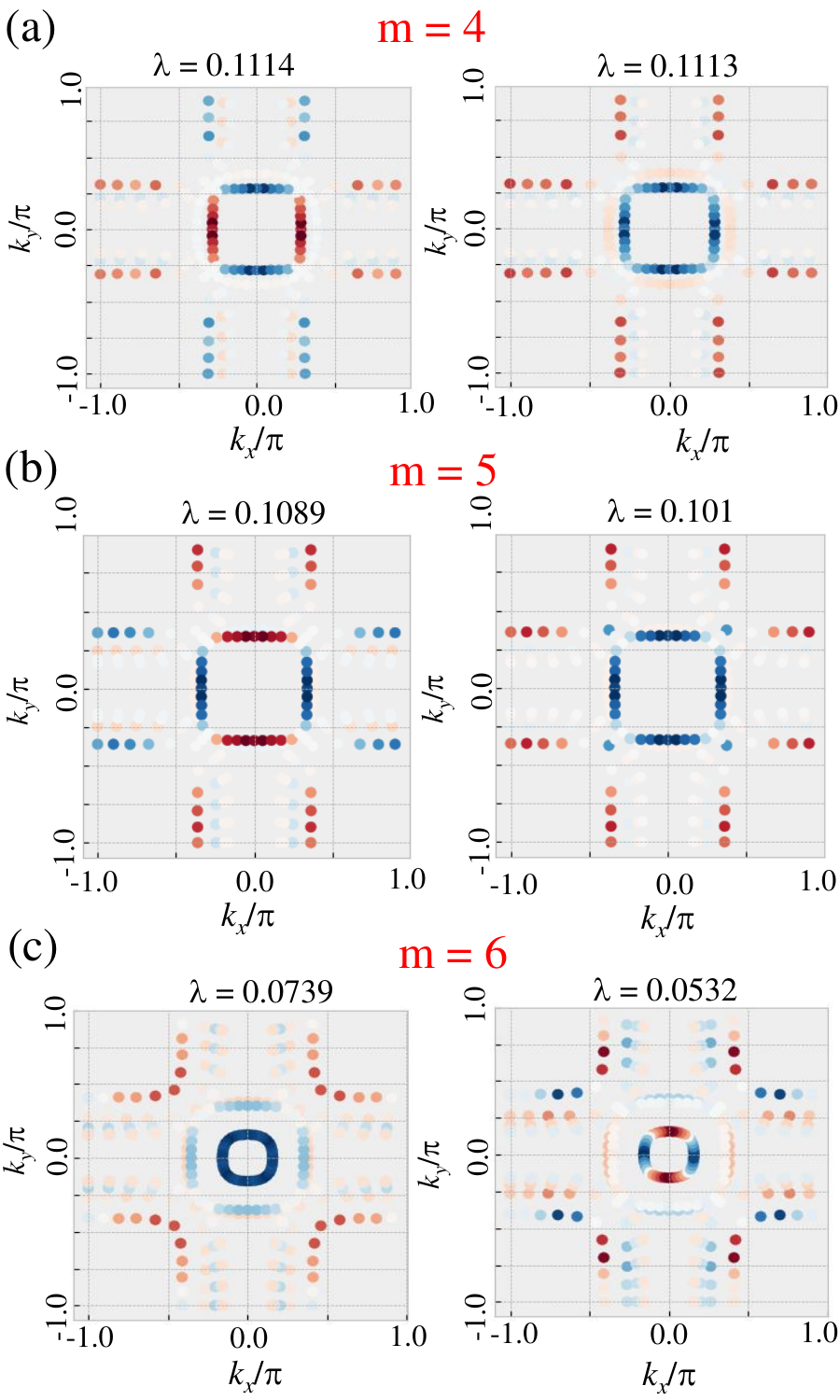}
\caption{The RPA calculated leading superconducting singlet gap structures $g_\alpha({\bf k})$ for momenta ${\bf k}$ on the Fermi surfaces for (a) La$_5$Ni$_4$O$_{13}$, (b) La$_6$Ni$_5$O$_{16}$, and (c) La$_7$Ni$_6$O$_{19}$ with corresponding pairing strengths $\lambda$ at 30 GPa. The sign of $g_\alpha({\bf k})$ is indicated by the color (red = positive, blue = negative), and its amplitude by the color darkness. We used Coulomb parameters $U=1.4$ eV, $U'=U/2$, and $J=J'=U/4$. The calculation was performed at $T=0.01$ eV.}
\label{Pairing}
\end{figure}

Figure~\ref{Pairing} shows our numerical results for the leading pairing instability in the high-order stacking La$_{m+1}$Ni$_m$O$_{3m+1}$ ($m = 4$ to 6), obtained by solving the eigenvalue problem in Eq.~(\ref{eq:pp}). Here, we used Coulomb parameters $U=1.4$ eV, $U'=U/2$, and $J=J'=U/4$, and the calculation was performed at a temperature $T=0.01$ eV.

For the four-layer $m = 4$ case, we find nearly degenerate $d_{x^2-y^2}$-wave and $s^\pm$-wave leading states with almost identical eigenvalues $\lambda_d = 0.1114$ and $\lambda_s = 0.1113$, respectively. For both the $d_{x^2-y^2}$-wave and the $s^\pm$ states, the gap is largest on the inner $\delta$ pocket with mainly $d_{3z^2-r^2}$ character and the $\beta$ sheets with mixed $e_g$ character  (see Fig.~\ref{Pairing}(a)).

For the $s^\pm$-wave state,  the phase of the gap changes signs between these two sheets. The five-layer $m = 5$ case also has nearly degenerate $d_{x^2-y^2}$-wave and $s^\pm$-wave leading solutions with similar  pairing strengths , $\lambda_d = 0.1089$ and $\lambda_s = 0.101$, respectively. In contrast to the 4-layer case, here the gap is largest on the inner $\alpha$ sheet with mixed $e_g$ character and the $\gamma_5$ sheet with $d_{3z^2-r^2}$ character. The sign changes in or between these two sheets, as displayed in Fig.~\ref{Pairing}(b). The six-layer $m = 6$ case shown in Fig.~\ref{Pairing}(c) has a slightly reduced $\lambda$ for the leading $s^\pm$-wave state and an even lower $\lambda$ for the $d_{x^2-y^2}$-wave channel. For both these states, the gap is largest on the inner $\delta$ pocket and the inner $\gamma_6$ sheet.

We note that the leading pairing strength $\lambda_d=0.1114$ for four-layer stacking La$_5$Ni$_4$O$_{13}$ at $U = 1.4$ eV is already much smaller than the pairing strength $\lambda_s=0.202$ we obtained for the trilayer system La$_4$Ni$_3$O$_{10}$ at 30 GPa but with a smaller interaction strength $U = 0.95$ eV \cite{Zhang:nc24}, suggesting that the pairing correlations in the trilayer stacking system are much stronger. Moreover, we found previously that the trilayer has smaller pairing strength than the bilayer 327-LNO at the same $U = 0.95$ ($\lambda_s\sim 0.39$). Since in our RPA treatment, the pairing strength $\lambda$ enters exponentially in the equation for $T_c$, i.e. $T_c=\omega_0e^{-1/\lambda}$ with a spin-fluctuation cut-off frequency $\omega_0$, this, and the new results in Fig.~\ref{Pairing} suggests that $T_c$ is expected to decrease in stoichiometric bulk systems as the layer stacking $m$ increases.

\subsection{D. Magnetic correlations}
Finally, we discuss the magnetic correlations in the RP nickelates La$_{m+1}$Ni$_m$O$_{3m+1}$ with different layer blocks $m$. To this end, we analyze the RPA enhanced spin susceptibility tensor $\chi({\bf q}, \omega=0)$ that is obtained from the Lindhart function tensor $\chi_0({\bf q})$ as
\begin{eqnarray}
\chi({\bf q}) = \chi_0({\bf q})[1-{\cal U}\chi_0({\bf q})]^{-1},
\end{eqnarray}
where all the quantities are rank-four tensors in the orbital indices $\ell_1, \ell_2, \ell_3, \ell_4$ and ${\cal U}$ is a tensor involving the interaction parameters~\cite{Graser2009}. The physical spin susceptibility is obtained by summing the pairwise diagonal tensor $\chi_{\ell_1\ell_1\ell_2\ell_2}({\bf q})$ over $\ell_1$, $\ell_2$.

\begin{figure}
\centering
\includegraphics[width=0.48\textwidth]{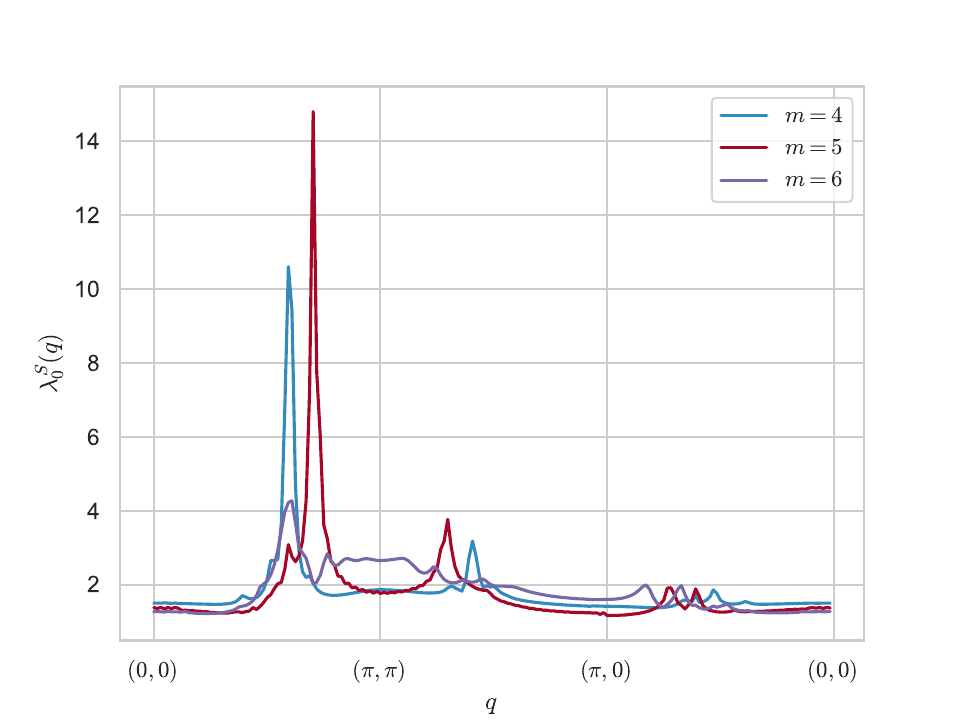}
\caption{ The RPA calculated leading eigenvector $\lambda^S_0(q)$ of the spin-susceptibility matrix $\chi_{\ell_1\ell_1\ell_2\ell_2}$ for La$_5$Ni$_4$O$_{13}$ ($m = 4$), La$_6$Ni$_5$O$_{16}$ ($m = 5$), and La$_7$Ni$_6$O$_{19}$ ($m = 6$) for the I4/mmm phase at 30 GPa.  The Coulomb parameters $U=1.4$ eV, $U'=U/2$, and $J=J'=U/4$, and the calculation was performed for a temperature of $T=0.01$ eV.}
\label{lambda_S_of_q}
\end{figure}

As displayed in Fig.~\ref{lambda_S_of_q}, the leading eigenvalue $\lambda^s_0({\bf q})$ of $\chi_{\ell_1\ell_1\ell_2\ell_2}({\bf q})$ for momenta ${\bf q}$ along a high-symmetry path in the Brillouin zone has the strongest in-plane peak at ${\bf q}=(0.6\pi, 0.6\pi)$, ${\bf q}=(0.7\pi, 0.7\pi)$, and ${\bf q}=(0.6\pi, 0.6\pi)$ for $m = 4$, 5, and 6, respectively. For the single-layer La$_2$NiO$_4$ case, the leading eigenvalue $\lambda^s_0({\bf q})$ has a maximum at ($\pi$, 0) or (0, $\pi$). This is different from the bilayer and trilayer systems~\cite{Zhang:nc24,Zhang:arxiv24}, for which we previously found that the in-plane peak  is near $(\pi, 0)$ or $(0, \pi)$ and $(\pi,\pi)$, respectively. In the RPA treatment, this difference arises from differences in the Fermi surface and the respective nesting wavevectors. This may be a consequence of the intraorbital and interorbital
hopping mechanisms with different electronic densities, as discussed in previous studies~\cite{Lin:prl21,Lin:cp,Lin:prb22}.

\begin{figure*}
\centering
\includegraphics[width=0.92\textwidth]{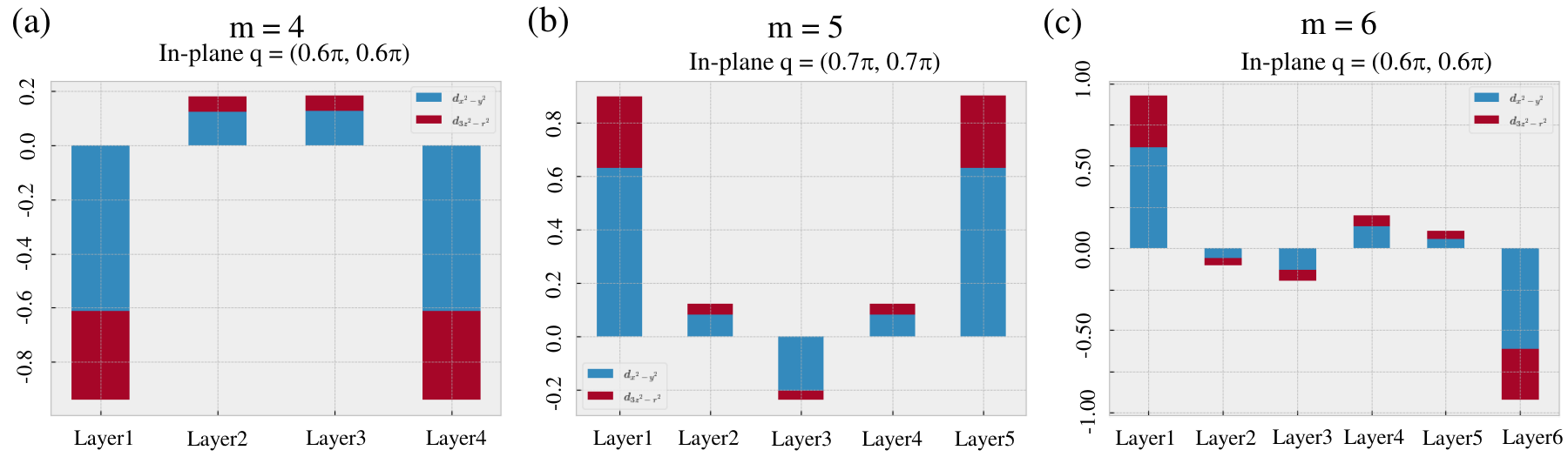}
\caption{ Leading eigenvector of the susceptibility matrix $\chi_{\ell_1\ell_1,\ell_2\ell_2}(q,\omega=0)$ at the in-plane $q$-vector for which its leading eigenvalue $\lambda^S_0(q)$ has a maximum, for (a) La$_5$Ni$_4$O$_{13}$, (b) La$_6$Ni$_5$O$_{16}$, and (c) La$_7$Ni$_6$O$_{19}$. The contributions of $d_{3z^2-r^2}$ and $d_{x^2-y^2}$ orbitals are shown by blue and red colors. The Coulomb parameters are $U=1.4$ eV, $U'=U/2$, and $J=J'=U/4$, and the calculation was performed for temperature $T=0.01$ eV.}
\label{Magnetism}
\end{figure*}

Figure~\ref{Magnetism} shows the corresponding eigenvector $\varphi^s_0({\bf q})$ at the in-plane ${\bf q}$ for which its eigenvalue $\lambda^S_0({\bf q})$ has a maximum.  One sees that the main contributions to the magnetic correlations come from the $d_{3z^2-r^2}$ orbitals from outer Ni layers. In addition, the out-of-plane, inter-layer magnetic correlations determined by the relative sign of the eigenvector contributions also strongly depend on the number of layers $m$. In the 4- and 5-layer systems, the outer Ni layers are coupled ferromagnetically, but in the 6-layer system, they have antiferromagnetic coupling. For La$_5$Ni$_4$O$_{13}$, the four layers are coupled as $\downarrow$\,--\,$\uparrow$\,--\,$\uparrow$\,--\,$\downarrow$ and the five layers are coupled as $\uparrow$\,--\,$\uparrow$\,--\,$\downarrow$\,--\,$\uparrow$\,--\,$\uparrow$ along the out-of-plane directions, while the La$_7$Ni$_6$O$_{19}$  has an $\uparrow$\,--\,$\downarrow$\,--\,$\downarrow$\,--\,$\uparrow$\,--\,$\uparrow$\,--\,$\downarrow$ coupling. In contrast to the trilayer case, we do not find a spin-zero Ni layer in high-order La$_{m+1}$Ni$_m$O$_{3m+1}$ with $m = 4$ to 6.

\section{IV. Conclusions and Discussions}
In this work, studying the high-symmetry I4/mmm phase under pressure, we systematically explored physical trends in the RP $m$-layer stacking nickelates La$_{m+1}$Ni$_m$O$_{3m+1}$.
Near the Fermi surface, the states are mainly contributed by Ni $3d$ orbitals, because O $2p$ orbitals are located deeper in energy, suggesting a common picture in nickelate superconductors involving a large charge-transfer gap between $3d$ and $2p$ states. Furthermore, the $d_{3z^2-r^2}$ orbital exhibits the bonding-antibonding or the bonding-antibonding-nonbonding splitting characters, depending on whether $m$ is even or odd.
Due to the in-plane interorbital hopping between the $d_{3z^2-r^2}$ and $d_{x^2-y^2}$ orbitals, the electronic occupations of both orbitals is not an integer, leading to a ``self-doping'' situation, as observed in other RP nickelate superconductors. Furthermore, the ratio of the in-plane interorbital hopping between the $e_g$ orbitals and in-plane intraorbital hopping between $d_{x^2-y^2}$ orbitals is found to increase with $m$.

For the single-layer $m = 1$ case, we found a strong spin-density-wave instability around $(\pi, 0)$ or $(0, \pi)$ due to the perfect nesting between two nearly flat $\gamma_1$ pockets. For the bilayer $m = 2$ and trilayer $m = 3$ cases~\cite{Zhang:nc24,Zhang:arxiv24}, the spin-fluctuation-driven $s^\pm$-wave state is dominant in both cases. For $m = 4$ and $m = 5$ cases, we find two nearly degenerate $d_{x^2-y^2}$-wave and $s^\pm$-wave leading states. However, in La$_7$Ni$_6$O$_{19}$, this six-layer stacking system has a slightly reduced $\lambda$ for the leading $s^\pm$-wave state and an even lower $\lambda$ for the $d_{x^2-y^2}$-wave channel. In general, at the level of the RPA treatment, {\it the superconducting transition temperature $T_c$ decreases as the layer stacking number $m$ increases in stoichiometric bulk systems}.

Similar to the pairing instability, magnetic correlations are found to be quite complex in the RP nickelates, depending on the stacking layer number $m$.
As mentioned above, single-layer La$_2$NiO$_4$ has a strong stripe instability characterized by ${\bf q} = (\pi, 0)$ or $(0, \pi)$, similar to our previous finding on bilayer La$_3$Ni$_2$O$_7$~\cite{Zhang:nc24}.
For the trilayer La$_4$Ni$_3$O$_{10}$~\cite{Zhang:arxiv24}, the peak of the magnetic susceptibility was found at ${\bf q} = (\pi,\pi)$.
For La$_5$Ni$_4$O$_{13}$ ($m = 4$) and  La$_7$Ni$_6$O$_{19}$ ($m = 6$), the peak position is shifted to ${\bf q} \approx (0.6\pi, 0.6\pi)$, and for La$_6$Ni$_5$O$_{16}$ ($m = 5$) the peak position is at ${\bf q} \approx (0.7\pi, 0.7\pi)$.
Moreover, the out-of-plane magnetic correlations strongly depend on the number of layers $m$.

\begin{figure*}
\centering
\includegraphics[width=0.8\textwidth]{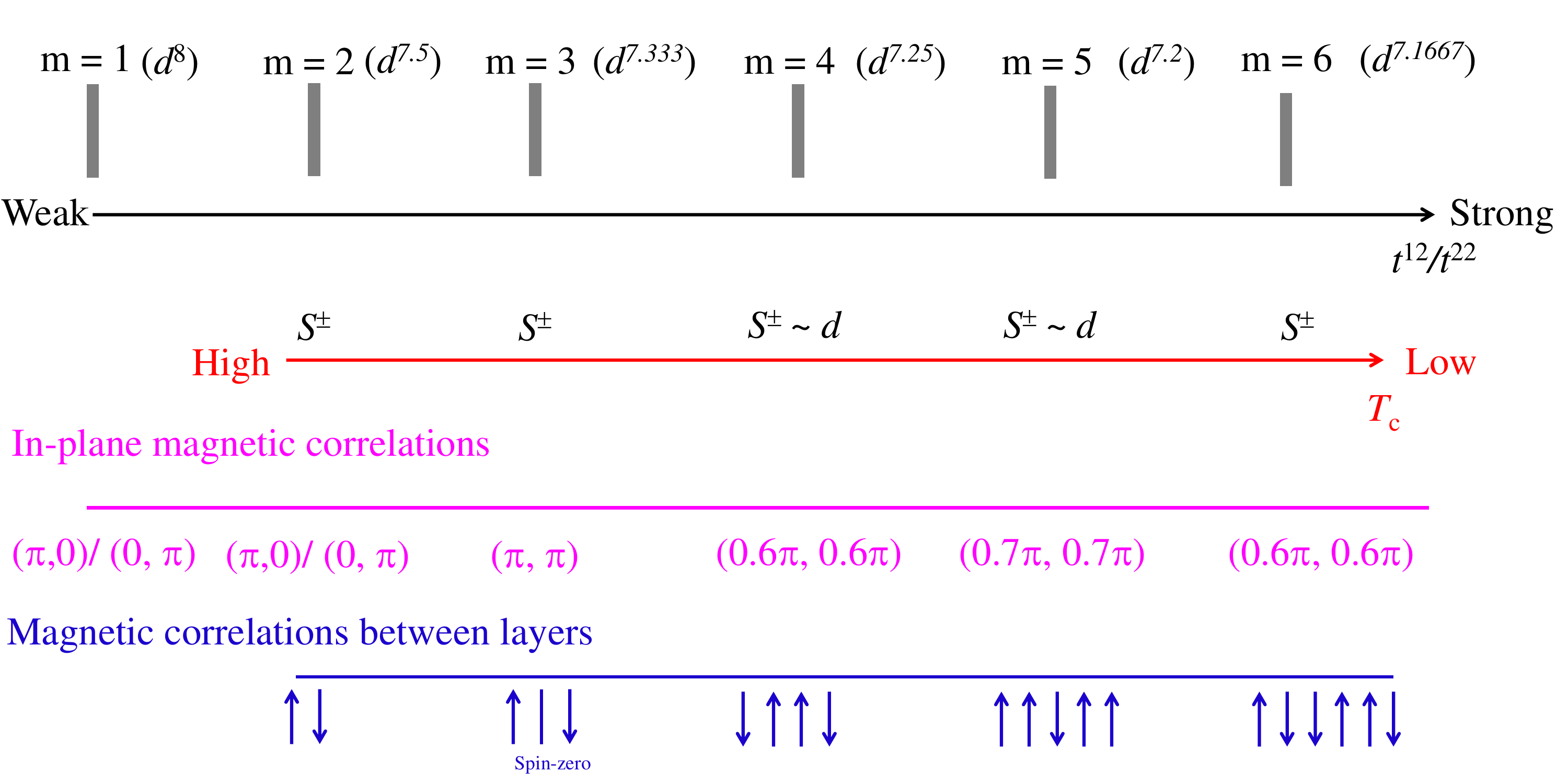}
\caption{ Summary of our main findings for the I4/mmm phase of the La$_{m+1}$Ni$_m$O$_{3m+1}$ family with different layer stacking number $m$ under pressure.}
\label{Summary}
\end{figure*}

Figure~\ref{Summary} summarizes our findings. We hope that our comprehensive study of the $m$ dependence of magnetic and pairing properties will stimulate experimental efforts to verify our theoretical predictions. The thin film of high-order La$_{m+1}$Ni$_m$O$_{3m+1}$ ($m = 4$ and $m = 5$)~\cite{Li:apl20} and Nd$_{m+1}$Ni$_m$O$_{3m+1}$ ($m = 4$ and $m = 5$)~\cite{Sun:prb21} has been stabilized in experiments using reactive molecular-beam epitaxy. Very recently, Y.-F. Zhao et. al.~\cite{Zhao:prb25} found that tensile strain can induce a small $\gamma$ pocket from the lowest bonding state of $d_{3z^2-r^2}$ orbitals near M points in the I4/mmm phase of four-layer stacking La$_5$Ni$_4$O$_{13}$ and five-layer stacking La$_6$Ni$_5$O$_{16}$, while a compressive strain can not induce this pocket for both cases, suggesting that the pressure may not induce the $\gamma$ pocket as well. In the previous context of bilayer and trilayer nickelates, this pocket is believed to be crucial for superconductivity. Thus, the next step is to investigate in the high-order stacking nickelates La$_{m+1}$Ni$_m$O$_{3m+1}$ ($m = 4$ to 6) the connection between superconductivity and the hole pocket from the lowest bonding state of $d_{3z^2-r^2}$ orbitals. Especially, to explore if the superconductivity can be enhanced in those high-order La-based systems or in other rare-earth RP nickelates ($m = 4-6$ case) when the small $\gamma$ pocket from the lowest bonding state of $d_{3z^2-r^2}$ orbitals exists near the M point.

\section{Acknowledgments}
This work was supported by the U.S. Department of Energy, Office of Science, Basic Energy Sciences, Materials Sciences and Engineering Division.

\section{Data Availability}
The input parameters for the kinetic energy term in our calculations are available for reproduction of our results in a separate file in the Supplemental Materials. Any additional data are also available from the authors upon reasonable request.

\end{document}